\documentstyle[12pt]{article}
\begin{document}
\newcommand{\LS}{{L^{(S)}}}
\newcommand{\LD}{{L^{(D)}}}
\hoffset 0.5cm
\voffset -0.4cm
\evensidemargin 0.0in
\oddsidemargin 0.0in
\topmargin -0.0in
\textwidth 6.7in
\textheight 8.7in

\begin{titlepage}

\begin{flushright}
PUPT-1673 \\
hep-th/9612087\\
December 1996
\end{flushright}

\vskip 0.2truecm

\begin{center}
{\large {\bf Supersymmetry and  Membrane Interactions in M(atrix) Theory}}
\end{center}

\vskip 0.4cm

\begin{center}
{Gilad Lifschytz and Samir D. Mathur${}^*$}
\vskip 0.2cm
{\it Department of Physics,
     Joseph Henry Laboratories,\\
     Princeton University, \\
     Princeton, NJ 08544, USA.\\ 
    e-mail: Gilad, Mathur@puhep1.princeton.edu }

\end{center}

\vskip 0.8cm

\noindent {\bf Abstract} 

We calculate the potential between various configurations of membranes
and gravitons in M(atrix) theory. The computed potentials agree
with the short distance potentials between 
corresponding 2-branes and 0-brane configurations in 
type IIA string theory, bound to a large number of 0-branes
to account for the boost to the infinite momentum frame. 
We show that, due to the large boost, these type IIA configurations 
are almost supersymmetric,  
so that the short and long distance potentials actually agree.
Thus the M(atrix) theory is able to reproduce correct long
distance behavior in these cases.

\noindent                
\vskip 6 cm
\begin{flushleft}
${}^*$On leave from The Center for Theoretical Physics,
M.I.T., Cambridge, MA 02139
\end{flushleft}

\end{titlepage}

\section{Introduction}

Recently there has been a renewed effort towards understanding the
quantum structure of M theory \cite{wit1,ht}, which is expected 
to underlie all the
string theories that we know. One of the tools used is the kinematic
improvements obtained by using an infinite momentum frame to view
the excitations of the theory. It has been conjectured that in such a frame
the dynamics can be captured by the Yang-Mills theory \cite{wit}
that is obtained
for 0-branes at short distances and low velocities. The gauge group
would however be $SU(\infty)$, corresponding to the limit
of an infinite number of 0-branes \cite{bfss}. This Yang-Mills theory 
reproduces the action of the membrane \cite{mem} of  11-dimensional 
supergravity, which was one of the principal motivations for the conjecture.
Recently it has been shown that the membrane constructed in this fashion
exhibits the correct Berry's phase when transported around a 5-brane of the
theory \cite{berdoug}.

If this conjecture can be verified 
then it would be a candidate for a nonperturbative formulation
of M theory, and thus of string theory. Large string coupling corresponds
to a large radius for the compactification radius of the coordinate $X^{11}$.
When some of the directions of space are compact the $SU(\infty)$ Yang-Mills
quantum mechanics is to be replaced by an  $SU(\infty)$ Yang-Mills field theory
in those directions \cite{bfss,sus,tay,grt}. Another formulation of
the conjecture can be found in \cite{per}.

In this paper we explore some of the consequences of the
infinite momentum frame approach and the use of the Yang-Mills action
to explore the amplitudes in M theory. We do the following:

(a)\quad We examine the effect of using a large boost in the $X^{11}$
direction of M theory, as seen from the viewpoint of type IIA theory.
The latter is the compactification of M theory on a circle in the direction
$X^{11}$, and since we wish to compare to results in weakly coupled
type IIA, we take the radius of this circle to be small. Suppose we
wish to derive the
potential between a 2-brane and an anti-2-brane at large distance. This
configuration is not supersymmetric. But now suppose we boost
the system in the direction $X^{11}$. From a type IIA string theory
point of view, this amounts to adding a number $N$ of
zero branes to each 2-brane. We show  that
this resulting configuration becomes close to being supersymmetric,
as the number $N$ becomes large. We do
this by mapping the configuration by a sequence of dualities to
one where we have two clusters of zero branes, moving at a slow relative
velocity. This latter system is known to be close to supersymmetric,
and this is manifested in the fact that the interaction potential
is the same at small  and at large distances. The small distance
potential is computed by using only the ground states of the open strings
stretching between the clusters, while the long distance calculation used 
all the oscillator modes of the open strings. The miracle of supersymmetry
is that the contribution of the higher
oscillator modes cancels out for the low velocity case. Since the boosted 
2-brane - anti-2-brane system is related by exact dualities to the low velocity
zero brane system, we see that the miracle of supersymmetry can work
for the former system as well. Once we have the potential in the boosted frame,
we use simple kinematics to find the potential of the 2-brane - anti-2-brane 
system at rest (i.e. no boost). We obtain exactly the correct known result for
the potential at distances larger than string length.

(b)\quad We find the potential  between different branes 
\cite{dlp,gr} in M(atrix) theory
using the matrix technique postulated in \cite{bfss}.
It is useful to think of the matrix technique in the following way.
The 2-brane bound to a number $N$  of 0-branes can
be described in two equivalent ways. 

(1) As a 2-brane carrying a U(1) 
magnetic field of $N$
units.  

(2) As a SU(N) gauge theory defined with gauge fields and fundamentals
in a twisted bundle, with the twist corresponding to the presence of 
the 2-brane. This latter approach was studied in \cite{grt}.

We note that calculations of matrix theory correspond to using
the latter language, and the usual string calculation
corresponds to the former.
We carry out calculations
for the potential between (1) a  zero brane and a 2-brane, with relative
transverse
velocity,  
(2) a 2-brane 
and an anti-2-brane, (3) two 2-branes moving with a relative transverse 
velocity, (4) two orthogonal branes moving with a transverse velocity.
In all case we show that the above two approaches agree.
Due to the large amount of zero-branes bound to the membrane
these configuration become close to supersymmetric, and one expects
to find that the short distance potential agrees
with the leading long distance potential. We show
that this is indeed the case.
Thus the M(atrix) theory reproduces the correct results at long distance.   

The plan of this paper is as follows. Section 2 explores the algebra
of considering boosts in the $X^{11}$ direction and regarding them as
creating 0-branes in the type IIA theory, and shows that
the brane configurations become more supersymmetric as the
boost parameter is increased. In Section 3 we discuss in more
details the effects of boosting in a compact direction, and in particular 
Lorentz invariance.
In section 4 we give a detailed example of computing the long distance
potential between a two-brane and an anti-two-membrane, using large boosts and
short distance information. Section 5 discusses the examples
of potentials between various branes in M(atrix) theory. 
Section 6 is a general discussion.

While we were writing up this work, a paper appeared  
\cite{ab} with many similar results.

\section{The relation between boosts and supersymmetry.}

In this section we show that if we boost a 
non-supersymmetric configuration of branes then it becomes close
to supersymmetric.

Suppose we have the following question. We have a 2-brane, wrapped
on a 2-torus. We also have, at a large  transverse distance, an anti-2-brane,
which is just the 2-brane wrapped with the opposite orientation. We
wish to know what is the leading potential at long distances between
this pair of objects. Let us denote the potential as $V_2(r)$. 

This calculation can be done through computing the 1-loop amplitude
of open strings that start on one brane and end at the other.
Since we are at long distances, which are in particular longer
than string length, we expect that we will have to use all the oscillator
modes of the open strings. Another way to say this is that since this
configuration of branes breaks all supersymmetry, we do not have
the miracle that happens for instance with the force between two
zero branes at low relative velocity: the long distance force law calculation
gives the same result as the short distance force law calculation, with the
latter using only the ground states of the open strings stretched between
the zero branes. But we shall see that after the boost in the
direction $X^{11}$ the system does indeed become close
to supersymmetric, so that the long distance potential can be found from the
lowest modes of the open string.

\subsection{The duality transformations.}

We proceed to calculate the potential between the 2-brane  and the anti-2-brane
 by
a sequence of dualities, which will reduce the problem to the force
between clusters of zero branes at low relative velocity. For the
system of the 2-brane and anti-2-brane at rest, the interaction potential
is known to be
\begin{equation}
V_2=\frac{16 \Gamma(5/2)(\LS)^{2} L_1 L_2}{2\sqrt{2}(2\pi)^{5/2} b^5}
\label{qtwo}
\end{equation}
at distances large compared to the string length. Our goal is to reproduce
this result using only the knowledge of potentials that
can be computed using only the ground state of the open string (for
example the potential between slowly moving 0-branes).

(a)\quad We assume that type IIA theory is just the compactification on
$S^1$ of a Lorentz invariant 11-dimensional M theory. Instead of computing
the potential between the branes at rest, we imagine that each brane is 
travelling
at a speed close to that of light in the direction $X^{11}$. Thus each brane
has a momentum
\begin{equation}p_{11}= \frac{N}{R}
\end{equation}
in the direction $X^{11}$. Here the length of the $X^{11}$ direction is
\begin{equation}
L_{11}=2\pi R = gL^{(S)}
\end{equation}
$\LS$ is the length scale of the elementary string, which we will use
as a reference length throughout this calculation. (The elementary string
tension is $T^{(S)}=2\pi\LS^{-2}$.) Under a T-duality, a circle
of length $L=A\LS$ changes to a circle of length $L'=A^{-1}\LS$. Under
S-duality the coupling $g$ changes to
$g'=g^{-1}$ and a  length $L=A\LS$ changes to a length $L'=A\LD=Ag'^{1/2}\LS$.
($\LD$ is the length scale of the D-string.)

{}From the perspective of type IIA theory we
have a 2-brane bound to $N$ 0-branes, and an anti-2-brane bound to $N$
zero branes. (The number of 0-branes is the same on both the
2-branes because they had the same mass before boosting, and thus at the
same boost velocity will have the same amount of momentum.)
Let the transverse distance between them be $b=A\LS$.

At this stage, the coupling constant of the type IIA theory is $g_1$, say.
The sides of the torus on which the 2-brane is wrapped are
\begin{eqnarray}
L_1^{(1)}=B_1\LS \\
L_2^{(1)}=B_2\LS
\end{eqnarray}
The third compactified direction  and the transverse separation between the branes are
\begin{eqnarray}
L_3^{(1)}=B_3\LS \\
b_1=A\LS 
\end{eqnarray}

(b)\quad We T-dualise in one of the directions of the torus on which
the 2-brane is wrapped, say the direction 2 (which has  length $L_2$).
 The 2-brane then becomes a D-string, and the 
anti-2-brane becomes a D-string wrapped with the opposite orientation. The zero
branes become D-strings, wrapped in a direction perpendicular to the above
D-strings, with winding number $N$. For both the branes the direction
of winding of this latter winding is the same. The bound state of the 2-branes
with the zero branes thus becomes a D-string wrapped on a cycle of the torus
that is $(1,N)$ for the 2-brane-zero brane bound state and $(-1,N)$ for
the anti-2-brane - zero brane bound state. Note that at this stage the
D-strings are close to being parallel, if $N$ is large. We
know that this situation is close to being supersymmetric, since the 
case where the two D-strings are parallel is exactly supersymmetric, 
and the breaking of supersymmetry increases continuously with
the angle between the D-strings.  A consequence is that there will
be an agreement between the potential computed at long distances
and the potential computed at short distances for our present system.

\section{The issue of Lorentz invariance.}

In the above we have boosted the branes in the direction $X^{11}$.
This was a compact direction though, not a noncompact one, so one may wonder what are the Lorentz kinematics of such a boost. One of the ideas of the M(atrix) approach is that we take the size of the compact direction to be large, so that we obtain an 11-dimensional M-theory rather than the dimensionally reduced 10 dimensional type IIA string description. Of course the larger the compact circle, the more the number of units of momentum that we wish to have, in order to approach the large momentum frame.

For a large size of the compact direction, if we have  small momentum in that direction, we will of course approximate well the usual Lorentz kinematics obtained for non-compact spacetimes.
But what happens if the boost is large?
We argue below that once we boost to a momentum that will help us to enhance supersymmetry in the way referred to above, the Lorentz kinematics may not resemble that for a noncompact direction at all.

We argue as follows. The 2-brane and anti 2-brane were separated
in a direction transverse to $X^{11}$, and if the Lorentz kinematics of noncompact space were to be a good approximation, then distances
transverse to the boost direction do not shrink or expand. But in this system of 2-brane - anti-2-brane, we know that in the absence of any boost, there is a tachyonic instability, when the transverse separation becomes equal to string length. This instability can be seen in a divergence of the 1-loop potential between the branes, or more dynamically in a divergent phase shift in a scattering process, when the impact parameter approaches string distance.

If the Lorentz kinematics of non-compact space were to hold, then we would expect the tachyonic instability distance to not change when the boost in the direction $X^{11}$ were applied. But as we saw above, by a T-duality the boosted system of branes can be mapped to a system of D-strings that are almost parallel.
(The noncompact directions were not affected in this change.)
But with D-strings, we know that there is a tachyonic instability at a transverse separation of string length only when they are antiparallel; if we bring the relative orientation towards that of parallel D-strings then the tachyonic instability distance becomes shorter, and in fact becomes zero when the D-strings are exactly parallel. 

Since the case at hand has mapped to almost parallel D-strings, we do have a big reduction in the tachyonic instability distance, when we compare the boosted and non-boosted cases. Note that this occurence of  almost parallel D-strings was exactly what helped us get the approach to supersymmetry, which as we will see below, allows us to use the Yang-Mills approximation to the full open string theory that governs that long distance interaction between the branes. Thus we conclude that if we boost enough to get the needed supersymmtry, then we cannot ignore the compactness of the direction $X^{11}$ in the kinematics of the boost.

There are potential counters to the argument above; we do not know really the strong coupling physics of type IIA, so we do not know much about what happens for large $X^{11}$. The tachyonic instability argument is a 1-loop effect of open strings, and it may well be that other loops somehow contrive to cancel the change in tachyonic instability seen at this lowest order. But in any case
this would be something that needs to be demonstrated; the limits of large radius for $X^{11}$ and large momentum in direction $X^{11}$ are not easily interchangeable.

\section{Mapping to the 0-brane problem.}

Thus we have shown the required approach to supersymmetry, but 
we continue the sequence of dualities
 to map the problem to the the problem of slowly moving 0-branes, for which the
potential was computed in \cite{dkps}.

At this stage the coupling is
\begin{equation}
g_2=g_1B_2^{-1}
\end{equation}
The lengths of the sides of the torus are
\begin{eqnarray}
L_1^{(2)}=B_1\LS \\
L_2^{(2)}=B_2^{-1}\LS 
\end{eqnarray}
The third compactified direction and the transverse separation are still
\begin{eqnarray}
L_3^{(2)}=B_3\LS  \\
b_2=A\LS
\end{eqnarray}

(c)\quad We S-dualise this configuration to get elementary strings instead
of D-strings in the same configuration as above. We have
\begin{eqnarray}
g_3=g_1^{-1}B_2 \\
L_1^{(3)}=B_1\LD=B_1g_3^{1/2}\LS=B_1g_1^{-1/2}B_2^{1/2}\LS \\
L_2^{(3)}=B_2^{-1}\LD=B_2^{-1/2}g_1^{-1/2}\LS \\
L_3^{(3)}=B_3g_1^{-1/2}B_2^{1/2}\LS \\
b_3=A\LD=Ag_1^{-1/2}B_2^{1/2}\LS
\end{eqnarray}

(d)\quad We T-dualise in the direction where the winding numbers were
$\pm 1$, which is the direction 1.
 This converts the winding into momenta, so we have an $N$ times
wound elementary
string moving with one quantum of momentum transverse to itself on the torus, 
and a similar $N$ times wound elementary string moving with one unit of
momentum in the opposite direction, at the location of the other brane.
We have
\begin{eqnarray}
g_4=g_3[B_1g_1^{-1/2}B_2^{1/2}]^{-1}=g_1^{-1/2}B_1^{-1}B_2^{1/2} \\
L_1^{(4)}=B_1^{-1}g_1^{1/2}B_2^{-1/2}\LS \\
L_2^{(4)}=B_2^{-1/2}g_1^{-1/2}\LS  \\
L_3^{(4)}=B_3g_1^{-1/2}B_2^{1/2}\LS \\
b_4=Ag_1^{-1/2}B_2^{1/2}\LS
\end{eqnarray}

(e)\quad We T-dualise in the direction 3, obtaining a type IIB theory
\begin{eqnarray}
g_5=B_1^{-1}B_3^{-1}\\
L_1^{(5)}=B_1^{-1}g_1^{1/2}B_2^{-1/2}\LS  \\
L_2^{(5)}=B_2^{-1/2}g_1^{-1/2}\LS \\
L_3^{(5)}=B_3^{-1}g_1^{1/2}B_2^{-1/2}\LS \\
b_5=Ag_1^{-1/2}B_2^{1/2}\LS
\end{eqnarray}

(f)\quad We perform an S-duality. We are still in type IIB theory.
We have
\begin{eqnarray}
g_6=B_1B_3 \\
L_1^{(6)}=B_1^{-1/2}B_3^{1/2}B_2^{-1/2}g_1^{1/2}\LS \\
L_2^{(6)}=B_2^{-1/2}B_1^{1/2}B_3^{1/2}g_1^{-1/2}\LS \\
L_3^{(6)}=B_1^{1/2}B_3^{-1/2}B_2^{-1/2}g_1^{1/2}\LS  \\
b_6=AB_1^{1/2}B_2^{1/2}B_3^{1/2}g_1^{-1/2}\LS
\end{eqnarray}

(g)\quad We perform a T-duality in the direction 2. Then we have
\begin{eqnarray}
g_7=B_1^{1/2}B_2^{1/2}B_3^{1/2}g_1^{1/2} \\
L_1^{(7)}=B_1^{-1/2}B_3^{1/2}B_2^{-1/2}g_1^{1/2}\LS \\
L_2^{(7)}=B_2^{1/2}B_1^{-1/2}B_3^{-1/2}g_1^{1/2}\LS \\
L_3^{(7)}=B_1^{1/2}B_3^{-1/2}B_2^{-1/2}g_1^{1/2}\LS  \\
b_7=AB_1^{1/2}B_2^{1/2}B_3^{1/2}g_1^{-1/2}\LS
\end{eqnarray}

We have now reduced the problem to that of slowly moving
clusters of 0-branes at long distance.

 The fact that the clusters are
moving slowly follows from the fact that $N$ branes share one unit
of momentum, so that
\begin{equation}
v\approx {2\pi \over NL_1^{(7)}T_0}
\label{twonew}
\end{equation}
where $T_0=2\pi\LS g_7^{-1}$ is the mass of the zero brane ($N$ is large).

\subsection{Computing the potential.}

\bigskip
{\it 4.1.1 \quad The potential between the 0-brane clusters}
\bigskip

We now discuss the potential between the clusters of 0-branes. 
 If the spacetime were noncompact, the potential would be
\begin{equation}
V=\frac{N^2\Gamma(7/2)(2v)^4 (\LS)^{6}}{ \sqrt{2}(2\pi )^{7/2}
 r^7}
\end{equation}
where we have used that the relative velocity between the clusters is $2v$.

If the space was a 2-torus with sides $L_1, L_2$,
 then to find the potential we note that we have to
 consider winding modes of the open strings around the cycles of the
torus. An equivalent way to do this is to go to the
covering space of the torus and thus find that each
 0-brane at one location receives a contribution
to its potential from an entire array of 0-branes at
 the other transverse location. 

There are two cases 
where we can simplify further in the sense that
we can replace the lattice sum by an integral. One is
 when the transverse distance $b$ is much larger than the
 compactification radii of
the torus. The other is when the density
of 0-branes on the compact space is high. When there is
one compact direction for instance, then the 0-branes 
form a bound state of size $\sim g^{1/3}\LS$ when the compact direction is
larger than this scale, but they behave as if they are spread out
uniformly on the circle if the compact direction is much smaller than
this scale \cite{mathur}. We assume that the same phenomenon holds
when we compactify on a 2-dimensional torus. We have the torus with an area
 which is much smaller than the
scale of the bound state of zero branes for small $g_1$. Thus we will have
the zero branes `smeared' over the torus, and we will naturally get
an integral over the position of the zero branes.

The lattice sum converted to an integral replaces the factor $b^{-7}$ obtained
in noncompact spacetime by
\begin{equation}
\int (b^2+z^2)^{-7/2}\frac{d^2z}{L_1 L_2}=( b^5 L_1 L_2)^{-1}
\frac{2\pi}{5} 
\end{equation}
where $L_1, L_2$ are the lengths of the compactified directions.

We will be interested in the case where one additional direction
has been compactified, and the force is computed at long distances. The effect of this latter compactification is the same on both the computation of the force between the 2-branes and on the computation of the force using 0-branes. In each case a potential $\sim b^{-5}$ is to be modified to
\begin{equation}
{1\over L_3}\int {dx\over (b^2+x^2)^{5/2}}={1\over L_3}{4\over 3b^{4}}
\end{equation}
where $L_3$ is the length of the additional compactified direction.

Thus for the clusters of 0-branes with small transverse velocities,
which we obtained by dualities from the initial problem concerning 2-branes,
we get the potential 
\begin{equation}
V_{zero}=\frac{N^2 \Gamma(7/2)(2v)^4 8\pi(\LS)^{6}}
{ 15  \sqrt{2}(2\pi )^{7/2} b^4 L_1L_2L_3}
\label{qthree}
\end{equation}
where now $L_1, L_2, L_3$ are the compactification lengths for the 0-brane case.

\bigskip
{\it 4.1.2\quad Potential between 2-branes}
\bigskip

Let us now ask the question: What is the relation between the potential
of the boosted 2-brane - anti-2-brane configuration and the 
potential of the same configuration without the
boost. If the boost were in a noncompact direction, then we would
have the following analysis.
 The energy of the
moving configuration is
\begin{equation}
E=[P_{11}^2+[2L_1 L_2 T^{(2)}+V_2]^2]^{1/2}
\approx P_{11}+\frac{2L_1 L_2 T^{(2)}V_2}{ P_{11}} +{2(L_1L_2T^{(2)})^2\over P_{11} }
\end{equation}
The middle term in the last expression is the potential energy that we attribute to the interaction between branes. Here $P_{11}=2N/R$ is the total momentum of the system (arising because of the 
boost),
the term $2L_1L_2T^{(2)}$ is the rest energy of the two branes when they
are separated by infinite distance, and
$V_2$ is the potential energy of interaction for the 2-brane anti-2-brane 
system at rest (i.e. without boost). Note that  
$V_2$ is proportional to $B_1B_2$. Thus we would
conclude that the interaction potential for the boosted configuration is
\begin{equation}
V_{\rm boost}=\frac{2L_1 L_2 T^{(2)}V_2}{ P_{11}} 
\end{equation}

Since the direction $X^{11}$ is actually compact, we again go to the covering 
space and note that each 2-brane receives a contribution to its potential
energy from a 1-dimensional array of anti-2-branes, all spaced a distance 
$2\pi R$ apart in the $X^{11}$ direction. We wish to compare our results to
a leading order in $g$ calculation of type IIA theory, so we assume
the $R=g\LS/2\pi$ is small, at least as compared to the 
transverse separation between the branes. Then the sum over the 1-dimensional
array can be replaced by an integral. (It is the result of this lattice 
sum/integral
which the potential $V_2$ describes for the case of the branes at rest.)

But if the branes are both moving rapidly in the direction $X^{11}$, then
we have the following effect. The separation between branes is $2\pi R$ in
the frame that we have been working with, but in the frame moving with
the 2-brane, the spacing of the anti-2-branes is dilated to $2\pi R/\gamma$
where $\gamma^{-1}=\sqrt{1-v_{11}^2}$. Note that for large boosts,
\begin{equation}
\gamma\approx p_{11}/M =\frac{N}{R L_1 L_2T^{(2)}}
\end{equation} 
($M$ is the mass of the 2-brane.)

Thus when we view the 2-brane - anti-2-brane system as having just a large
velocity in the $X^{11}$ direction, then we expect a potential energy
changing with separation as
\begin{equation}
V_{\rm boost}=\frac{2L_1 L_2 T^{(2)}V_2}{P_{11}\gamma} 
\label{qone}
\end{equation}
where $V_2$ is the potential between the 2-brane  anti-2-brane system at rest.

For large separations with just two directions compactified $V_2$ is known to be
\begin{equation}
V_2=\frac{16 \Gamma(5/2)(\LS)^{2} L_1 L_2}{2\sqrt{2}(2\pi)^{5/2} b^5}
\end{equation}
With one additional direction compactified, at large separation we have
\begin{equation}
V_2=\frac{32 \Gamma(5/2)(\LS)^{2} L_1 L_2}{3\sqrt{2}(2\pi)^{5/2} L_3 b^4}
\label{threenew}
\end{equation}

\bigskip
{\it 4.1.3\quad Comparing the 0-brane and the 2-brane potentials}
\bigskip

We have performed several dualities in mapping the problem of the 2-branes boosted along $X^{11}$ to the problem of slowly moving 0-branes. To see if the potentials in the two cases agree, we must
note the change of potential under duality. Let the potential between the boosted 2-branes be $V^{(1)}\LS^{-1}$, where $V^{(1)}$ is
a dimensionless number. Under a T-duality a potential $V\LS^{-1}$ goes to $V\LS^{-1}$, while under an S-duality a potential $V\LS^{-1}$ goes to $V\LD^{-1}=V\LS^{-1}g^{-1/2}$. Following these changes we find that the potential between the 0-branes which we get at the end of our dualities, which we call $V^{(7)}\LS^{-1}$, will be given through
\begin{equation}
V^{(7)}=V^{(1)}g_1^{1/2}B_1^{-1/2}B_2^{-1/2}B_3^{-1/2}
\label{onenew}
\end{equation}

For the 0-brane potential, in (\ref{qthree}) we have to use the velocity (\ref{twonew}). We have to use sides of the torus as $L_1^{(7)}, L_2^{(7)}, L_3^{(7)}$.  For the boosted 2-brane potential we have to use (\ref{qone}), with $V_2$ given by (\ref{threenew}) and 
with the sides of the torus being $L_1^{(1)}, L_2^{(1)}, L_3^{(1)}$.

Substituting these values we find the exact agreement, using (\ref{onenew}), between the potential \ref{qone} of the boosted 2-branes and the potential (\ref{qthree}) implied by the moving 0-branes:
\begin{equation}
V_{boost}\equiv V^{(1)}\LS^{-1}=g_1^{-1/2}B_1^{1/2}B_2^{1/2}B_3^{1/2}V^{(7)}\LS^{-1}\equiv V_{zero}
\end{equation}
where we have used that
\begin{equation}
V^{(1)}={(B_1B_2)^3\Gamma(5/2)32\over  B_3 N^2 3\sqrt{2} (2\pi)^{5/2} A^4}
\end{equation}
\begin{equation}
V^{(7)}={g_7^4\Gamma(5/2)32\LS^{11}\over  b_7^4[L_1^{(7)}]^5
L_2^{(7)}L_3^{(7)} N^2 3\sqrt{2} (2\pi)^{5/2} }
\end{equation}

 Thus we have shown that
we can recover the long distance force between the 2-brane and anti-2-brane, by making a boost in the $X^{11}$ direction, mapping to the problem of slowly moving 0-branes where short and long distance potentials agree, and boosting back.

\subsection{A comment on boosts in compact directions.}

Note that here we used relativistic kinematics in a domain where we
may not have expected it to be valid. We have taken a very small
size for the compact direction $X^{11}$, since we took small coupling
$g_1$; indeed this size is much smaller than the 11-dimensional Planck length.
One may therefore wonder what are the kinematic rules for relating the
potentials between frames boosted by large amounts in small compact
directions. One effect of the compactness we have already seen: the potential
is further scaled by a factor $\gamma^{-1}$ due to fact that we
have the potential coming from a sum over copies of the brane in
the periodic direction, and the separation of copies was dilated by the boost.

There is a related effect that we describe as follows. Consider the
potential between two 0-branes in type IIA theory. Let a pair of directions
be compact. If we boost in the direction $X^{11}$ then we get that each
0-brane is replaced by a large number $N$ of 0-branes. What is the
potential between such 0-branes for low transverse velocities?

We know that for zero branes
that are not boosted  the potential is $\sim v^4/b^7$ for $b<L_1, b<L_2$ and
is $\sim v^4/b^5$ for $b>L_1, b>L_2$. (Here $L_1, L_2$ are
the compactification lengths.)

Now consider the boosted case. For a large boost, when the number 
$N\alpha'/(L_1L_2)$ is sufficiently large, the 0-branes are smeared over the
torus, as discussed above. Then we see that whether we have
 $b<L_1, b<L_2$ or  $b>L_1, b>L_2$ we get the potential $\sim v^4/b^5$
in either case. 

Thus we see that if we examine a system of branes after adding a high boost
in a compact direction, then the potential that will result will agree
with the `long distance' potential between the branes. Here the term
`long distance' denotes the fact that if there are any compact directions in
the spacetime, the potential will have the fall off pertinent to separations
much larger than the compactification scale, and not to the potential at
distances smaller than the compactification scale.

\section{Potential between Branes}
In the following subsections 
we will calculate the phase shift of an object when it is 
scattered from another object, and
extract from it 
the potential between them. We will consider configurations of a graviton
scattering off a membrane , two parallel membranes moving with
a relative velocity, two
 orthogonal membranes moving with a relative velocity,
and  membrane
anti-membrane configuration (with no relative motion). 

These calculation will be done in the frame work of
BFSS \cite{bfss}, which correspond to using
only the ground states of the open strings between 0-branes.
It is therefore not surprising that the results agree with the
corresponding calculations carried out in the type IIA theory
when the brane separations are short compared to string length.
(We verify this agreement in several cases.)
What may appear more surprising is that there is agreement with 
type IIa calculations where the velocities are small, even when the
brane separation is large compared to the string length. But the reason for
this, as mentioned in the introduction, is that these calculations all
pertain to a situation where there is a large boost in the $X^{11}$ direction,
and this brings the interacting system close to being supersymmetric.
In such systems the potentials at long and short distances
agree \cite{dkps}.
The potentials at large distances are what would characterise the
interaction of membranes in the 11-dimensional supergravity theory
\cite{gil1,gil2}, and
it may even be expected that non-renormalisation theorems protect the
form of these potentials (for low velocities) as we go to
strong coupling. If these expectations were to hold, then the M(atrix)
calculations would tell us something about  M theory in the sense that
was hoped for in \cite{bfss}.

In the type IIA theory we are actually working at  small coupling. The M(atrix)
calculations are done at one loop, and thus also assume a small coupling.
Non-renormalisation theorems may however carry over the results to
larger coupling.

\subsection{The set up}
In this section we will set up the formalism for calculating the 
phase shift and the 
velocity dependent potential between two moving objects in M(atrix)
theory.
Let us start with the Lagrangian \cite{bfss,kp,dfs}, we take the string
length $l^{2}_{s}=2\pi$, the signature is $(-1,1 \ldots,1)$, and  
$D_t X=\partial_t X -i [A_0,X]$,
\begin{equation}
L=\frac{1}{2g}Tr\left[D_{t}X_{i}D_{t}X^{i}+2\theta^{T}D_{t}\theta-
\frac{1}{2}[ X^{i},X^{j}]^{2}-2\theta^{T}\gamma_{i}[\theta,X^{i}] \right].
\label{l}
\end{equation}
The supersymmetry transformations are
\begin{eqnarray}
\delta X^{i}=-2\epsilon^{T} \gamma^{i} \theta \nonumber\\
\delta \theta =\frac{1}{2}\left[ D_{t}X^{i}\gamma_{i}+\frac{1}{2}
[X^i, X^j]\gamma_{ij} \right]\epsilon +\epsilon' \nonumber \\
\delta A_{0} =-2\epsilon^{T}\theta
\label{susy}
\end{eqnarray}

We would like to calculate the phase shifts of a
graviton scattering off various 
configurations and of membranes scattered off 
various configurations. This can
be done by calculating the vacuum energy of the Lagrangian (\ref{l}) in
a background corresponding to the configuration of objects that we are
interested. This is done by giving expectation values to various matrices and
expanding perturbatively all quadratic term around that background.

As in \cite{dkps} we find it convenient to work 
in a background covariant gauge
 \footnote{We would like to thank Dan Kabat for a very
helpful discussion on this point}.
For that let us go slightly back to the Yang mills 
theory from which the above Lagrangian was derived.  Starting with
\begin{equation}
L_1=Tr\{-\frac{1}{4}F^{\mu\nu}F_{\mu\nu}-\frac{1}{2}(\bar{D}^{\mu}A_{\mu})^2
+L_g\}.
\label{l1}
\end{equation}
where $\bar{D}^{\mu}A_{\mu}=\partial^{\mu}A_{\mu}-i[B^{\mu},A_{\mu}]$,
$B_{\mu}$ is the expectation value of $A_{\mu}$ and $L_g$ is the 
corresponding ghost term.
 One can define as in \cite{kp,dfs}
the fields $F_{0i}=\partial_{0}X_{i}-i[A_0,X_i]$ and $F_{ij}=-i[X_i,X_j]$.
If one chooses the $B_{\mu}$ such that $B_0=0$ and the other $B_{i}$ solve the
equation of motion then we can expand (\ref{l1}) 
to quadratic order in the fluctuations
around the background fields and find ($X_i=B_i +Y_i$)
\begin{eqnarray}
L_{2} & = & \frac{1}{2}Tr \{ (\partial_0 Y_{i})^2-(\partial_0 A_{0})^2
-4i\dot{B}_i[A_0,Y^i]+\frac{1}{2}[B_i,Y_j]^2+\frac{1}{2}[B_j,Y_i]^2 \nonumber \\
  & + & [B_i,Y^j][Y^i,B^j]+[B_i,Y^i][B_j,Y^j]-
[A_0,B_i]^2+[B_i,B_j][Y^i,Y^j]  \nonumber \\
 & + & \partial_{0} C^* \partial_{0} C+[C^*,B^i][B_i,C] \}.
\label{l3}
\end{eqnarray}
This gauge is more convenient to work with as there will be only few
and simple off diagonal
terms in the mass matrix of the form $[B_i, B_j][Y_i,Y_j]$ and
$4\dot{B}_i[A_0,Y^i]$, all other off diagonal terms cancel.
To this Lagrangian
we must add the fermionic terms from equation (\ref{l}).

Now in order to compute scattering of two objects one has 
to insert the appropriate
expectation values and integrate out the massive fields. 
As each object by itself
will have zero vacuum energy one has only to consider
 fields that connect  the
two configurations, these will be the off diagonal matrix
 elements, which correspond
in the language of strings to the degrees of freedom of 
the virtual strings stretched
between the two objects.
So we will take the following form for $Y_i$ and $\theta$.
\[
Y_i=\left(
\begin{array}{cc}
0 & \phi_i \\ 
\varphi_i & 0 
\end{array}
\right), \ \
\theta=\left(
\begin{array}{cc}
0 & \psi \\ 
\chi & 0 
\end{array}
\right)
\]
Where the fermion are real sixteen component spinors, 
$\varphi=\phi^{\dagger}$, $\chi=\psi^{T}$ and these relationships 
should be understood in the matrix space.
To calculate the phase shift one must calculate the
 one vacuum energy of the 
off diagonal components of the matrices. This is done at
a one loop level by computing the
determinants of the operators $(\partial_{0}^{2} +M^2)$
where $M^2$ is the mass matrix squared of all the fields including of course
the fermions.


\subsection{Graviton Membrane scattering}

In this section we will compute the one loop phase shift for a graviton 
scattered off a membrane as a function of the 
distance between them and the graviton 
transverse velocity. The graviton will actually be represented by one
zero-brane with the understanding that to leading order
 we have to multiply the
answer by $N_1$, the number of zero-branes the graviton is made off. 
The membrane is wrapped on a very large torus.
>From the phase shift 
 we will extract the short distance behavior, the
static potential between them and the 
velocity dependent
 potential at long distances,

The background configuration is 
\[
B_8=\left(
\begin{array}{cc}
P &  0  \\ 
0 & 0 
\end{array}
\right),
B_9=\left(
\begin{array}{cc}
Q &  0  \\ 
0 & 0 
\end{array}
\right),
B_7=\left(
\begin{array}{cc}
bI &  0  \\ 
0 & 0 
\end{array}
\right),
B_1=\left(
\begin{array}{cc}
Ivt &  0  \\ 
0 & 0 
\end{array}
\right).
\]

$P,Q,I$ can be thought of as  matrices, [Q,P]=ic.
$\phi$ is a column vector while $\varphi$ is a row vector, and
one should think of them as living in the adjoint representation of the gauge
group.
We are going later to calculate the vacuum energy in this configuration.
For that we will have to calculate a large number  of determinants, because
the $P,Q,I$ matrices will be $N \times N$ matrices. we will first
diagonalise the mass square matrix $M^2$, in the space of
of $Y_i,A_0,C$ and then we will look at the spectrum in each one separately.

Plugging the $B$'s 
 into the Lagrangian (\ref{l3}) one finds that the mass term
($\varphi^i M^2 \phi_i$)  in the 
space of $(Y_2,\ldots ,Y_7,C)$ is proportional to the
 identity with the proportionality
constant being $2H$ and $H=P^2+Q^2+b^2+v^2t^2$ .

 In the space of $A_0,Y_1$ there are also off diagonal terms of $\pm 4iv$
\[
M^2_{A_0Y_1}=2\left(
\begin{array}{cc}
-H&  -2i v \\ 
2i v & H
\end{array}
\right)
\]
In the space of $Y_8,Y_9$ one has also off diagonal terms $\pm4ic$.
\[
M^2_{Y_8Y_9}=2\left(
\begin{array}{cc}
H&  -2 ic\\ 
2 ic & H
\end{array}
\right)
\]

The way these results are derived is just to compute the trace of the matrices
in equation (\ref{l3}). For example
\[
Tr[B_8,Y_i][B_9,Y_j]=-\left(
\begin{array}{cc}
P\phi^i\varphi^jQ &  0 \\ 
 0  & \varphi^iPQ\phi^j
\end{array}
\right)
\]
Then 

\begin{eqnarray}
Tr[B_8,Y_i][B_9,Y_j] & = &
-(P_{\alpha \beta}\phi_{\beta}^{i}\varphi^{j}_{\sigma}
Q_{\sigma \alpha}+\varphi^{i}_{\sigma}P_{\sigma \alpha}Q_{\alpha \beta}
\phi^{j}_{\beta}) \nonumber \\
& = & -\varphi^{j}_{\beta}(QP)_{\beta \alpha}\phi^{i}_{\alpha}
-\varphi^{i}_{\beta}(PQ)_{\beta \alpha}\phi^{j}_{\alpha}.
\end{eqnarray}

The ghost have the same mass $M^2=2H$ as some of the $Y$'s.
We will eventually calculate the determinant of the operator 
$(\partial^{2}_{0} +M^2)$ and then the ghost 
contribution will cancel some of the 
boson contribution.
The determinant calculations will be done in Euclidean space, where
$t=i\tau$ and $A_{0}=-iA_{\tau}$, so the mass matrix in the
$A_{\tau},Y_{1}$ will  be  (remembering that the $Y_{i}$ kinetic term changes
sign but that of $A_{0}$ does not)
\[
M^2_{A_{\tau} Y_1}=2\left(
\begin{array}{cc}
H&  -2 v \\ 
2 v & H
\end{array}
\right)
\]
Now in Euclidean space 
we will be left with four complex bosons with $M^2=2H$, one boson with
$M^2=2H-4c$, one with $M^2=2H+4c$, one with $M^2=2H+4iv$ and 
one with $M^2=2H-4iv$. These complex bosons when represented in
terms of real bosons will give twice the number of bosons but
with half the mass squared.

Let us now turn to the fermions described by the matrix 
valued sixteen real spinor
$\theta$. Inserting the $B$'s into the fermionic part on finds
(remembering the fermions are real)
\begin{equation}
m_f=\gamma_8 P + \gamma_9 Q + \gamma_7 b + \gamma_1 vt.
\label{mf02}
\end{equation}  
It is easier to convert the fermion determinants to the form
det$(\partial^{2}_{0} +M_{f}^{2})$, with
\begin{equation}
M_{f}^{2}=H-ic\gamma_8 \gamma_9+v\gamma_1
\label{Mf02}
\end{equation}
This can be diagonalised and we find: four fermions with
\footnote{In our conventions $\gamma^{2}_{i}=-1$} $M_{f}^{2}=H+c-iv$,
four with $M_{f}^{2}=H+c+iv$, four with $M_{f}^{2}=H-c-iv$ and
four with $M_{f}^{2}=H-c+iv$.

We need to find the eigenvalues of the operator $H$. As $[Q,P]=ic,$ 
The $Q,P$ operators can be represented as follows. On the space of $L_2$
functions of one variable $x$, $P$ acts as $-ic\partial_{x}$ and $Q$ acts
as $x$. Then the spectrum of $H$ is just the spectrum of an harmonic oscillator
with  the $n$'th eigenvalue being $H_n=b^2+v^2t^2+c(2n+1)$.
The eigenfunctions are just the usual ones of a one dimensional 
harmonic oscillator. They are the wave functions of the off diagonal 
elements of the matrices.

The determinants are
\begin{eqnarray}
{\det}^{-4}(-\partial_{\tau}^{2}+H_{n})
{\det}^{-1}(-\partial_{\tau}^{2}+H_{n}-2c)
{\det}^{-1}(-\partial_{\tau}^{2}+H_{n}+2c) \nonumber \\
{\det}^{-1}(-\partial_{\tau}^{2}+H_{n}-2iv)
{\det}^{-1}(-\partial_{\tau}^{2}+H_{n}+2iv) 
{\det}^{2}(-\partial_{\tau}^{2}+H_{n}+c+iv)\nonumber \\
{\det}^{2}(-\partial_{\tau}^{2}+H_{n}+c-iv)
{\det}^{2}(-\partial_{\tau}^{2}+H_{n}-c+iv)
{\det}^{2}(-\partial_{\tau}^{2}+H_{n}-c-iv) \nonumber
\end{eqnarray}

We can now calculate the phase shift for the graviton 
scattered off the membrane,
as in \cite{dkps} we use the proper time representation 
for the determinants.
define:
$\delta=\sum_{n} \delta_n$, 
$r^{2}_{n}=b^2+c(2n+1)$

 then we find 
\begin{equation}
\delta_n=\frac{1}{2}\int\frac{ds}{s}e^{-r_{n}^{2} s}\frac{1}{\sin sv}
(4+2\cosh 2cs+2\cos 2vs-8\cos vs \cosh cs)
\label{de02}
\end{equation}
This can be integrated to give
\begin{equation}
e^{i\delta_n}=\frac{\Gamma(\frac{ir_{n}^{2}}{2v}+\frac{1}{2}-\frac{ic}{v})
\Gamma(\frac{ir_{n}^{2}}{2v}+\frac{1}{2}+\frac{ic}{v})
\Gamma^{6}(\frac{ir_{n}^{2}}{2v}+\frac{1}{2})(r_{n}^{2}-iv)}{
\Gamma^2(\frac{ir_{n}^{2}}{2v}+1-\frac{ic}{2v})
\Gamma^2(\frac{ir_{n}^{2}}{2v}+1+\frac{ic}{2v})
\Gamma^2(\frac{ir_{n}^{2}}{2v}-\frac{ic}{2v})
\Gamma^2(\frac{ir_{n}^{2}}{2v}+\frac{ic}{2v})(r_{n}^{2}+iv)}
\label{edel02}
\end{equation}
and
\begin{equation}
|e^{i\delta_n}|^2=\frac{(\cosh\frac{\pi r_{n}^{2}}{v}-
\cosh\frac{\pi c}{v})^4}{\cosh^6 \frac{\pi r_{n}^{2}}{2v}
 (\cosh\frac{\pi r_{n}^{2}}{v}+\cosh\frac{\pi 2c}{v})}
\label{norm02}
\end{equation}
For later comparison it is convenient to notice that
\begin{equation}
2\sum_{n=0} e^{-cs(2n+1)} =\sinh^{-1} cs ,
\end{equation}
 then the expression
for the phase shift becomes
\begin{equation}
\delta=\frac{1}{2}\int\frac{ds}{s}e^{-b^{2} s}
\frac{4+2\cosh 2cs+2\cos 2vs-8\cos vs \cosh cs}{2\sinh cs \sin sv}
\label{del02}
\end{equation}

If we are interested in the potential it is defined through
\begin{equation}
\delta=-\int dt V(b^2+v^2t^2).
\end{equation}
If $b^2 \gg c$ 
we can expand
 equation (\ref{del02}) in powers of $s$.
to find a long range potential,
\begin{equation}
V_{gm}=-\frac{N_1\Gamma(5/2)}{4c\sqrt{\pi}}\frac{(c^4+2v^2c^2+v^4)}{b^{5}}.
\label{v02}
\end{equation}
where we have restored the factor of $N_1$.

Let us compare this to a string calculation of the phase shift of 
a zero-brane scattering
off a bound state of a membrane and many zero-branes.
>From the string calculation this is just having a constant magnetic field 
$F$ on the two brane \cite{doug}. Then the phase shift using
the technique in  \cite{pol,bac}, is \cite{gil2}
($\tan(\pi \epsilon) =  F$, $\tanh(\pi \nu)= v$,
$\Theta(\rho)=\Theta(\rho,is)$),
\begin{equation}
\delta_{IIA}=\frac{1}{2\pi}\int \frac{ds}{s} e^{-b^2 s} B \times J,
\label{a20}
\end{equation}
\begin{eqnarray}
B& = &\frac{1}{2}f_{1}^{-6}\Theta^{-1}_{4}(i\epsilon s)
\frac{\Theta_{1}' (0)}{\Theta_{1}(\nu s)}, \nonumber \\
J& = &\{ -f_{2}^{6}
\frac{\Theta_{2} (\nu s)}{\Theta_{2}(0)}
\Theta_{3}(i\epsilon s)+f_{3}^{6}\Theta_{2}(i\epsilon s)
\frac{\Theta_{3} (\nu s)}{\Theta_{3}(0)} \nonumber\\ 
 & + & if_{4}^{6} 
\frac{\Theta_{4} (\nu s)}{\Theta_{4}(0)}\Theta_{1}(i\epsilon s)\}.
\label{bj20}
\end{eqnarray}
Now there are many zero-branes on the membrane due to the fact that we 
originally were (almost) in the infinite momentum frame, and momentum in the
eleventh direction is like having many 
zero-branes (for the classical solutions
see \cite{rustsy,bm}),
so $\epsilon \sim 1/2$. Let us take 
$\epsilon=\frac{1}{2}-c'$ (we assume $c'$ is small), and insert that to 
equation (\ref{bj20}). Using the properties of the theta functions we find
that when $F$ is large we get,

\begin{eqnarray}
B& = &\frac{1}{2}f_{1}^{-6}(-i\Theta_{1})^{-1}(ic's)
\frac{\Theta_{1}' (0)}{\Theta_{1}(\nu s)}, \nonumber \\
J& = &\{ -f_{2}^{6}
\frac{\Theta_{2} (\nu s)}{\Theta_{2}(0)}
\Theta_{2}(ic` s)+f_{3}^{6}\Theta_{3}(ic' s)
\frac{\Theta_{3} (\nu s)}{\Theta_{3}(0)} \nonumber\\ 
 & - & f_{4}^{6} 
\frac{\Theta_{4} (\nu s)}{\Theta_{4}(0)}\Theta_{4}(ic' s)\}.
\label{bj20c}
\end{eqnarray}

One can compute the long range potential 
(that is expanding in powers of$s$) 
for the case of a zero-brane
scattering off a bound state of a two-brane with many zero-branes 
from equations (\ref{bj20c},\ref{a20}).
The long range potential can be then expanded 
 to leading order in $v$ and $c$ (assuming
they are small) to give,
\begin{equation}
V_{IIA}=-\frac{\Gamma(5/2) (2v^2(\pi c')^2+(\pi c')^4+v^4)}
{4\pi c'\sqrt{\pi}}b^{-5}.
\label{s02c}
\end{equation}
exactly like the result from matrix theory when one identifies $c=\pi c'$,
and multiplies by $N_1$.
Now the magnetic field strength integrated over the two-brane is
supposed to give $2\pi N$ where $N$ is the number of zero-branes 
on the two-brane, and also the size of the $P,Q$ matrix
of the membrane. Using $F=1/c$,
 we find $c=\frac{2\pi R_8 R_9}{N}$ where $R_8, R_9$ 
are the radiuses of the torus the membrane is wrapped on. Indeed for
large enough $N$ $c$ is very small.
 
This gives an explanation from the point of view of type IIA theory of the 
parameter $c$.
How is it that a configuration which is not supersymmetric in M-theory
when calculating the long range force with the use of only the
lightest open string mode reproduce the correct long range force.

Notice that equation (\ref{bj20c}) is actually equivalent up to an
overall pre-factor to a calculation involving 
two relatively moving two-branes with a small
magnetic field on one of them. As the non moving membrane configuration
with no magnetic field is supersymetric, the velocity and small magnetic field
only break the supersymmetry very softly. 
This is a situation when one expects that a long range potential could
be reproduced by using only the lightest open string modes.
In fact this is the case. The short distance behaviour of
the type IIA configuration for large $F$ can be extracted by
expanding equation (\ref{bj20c}) in the limit of
large $s$. Then ($a_1$ is just a number) 
\begin{equation}
B \times J =a_1\frac{4+2\cosh 2cs + 2\cos 2vs 
-8\cosh cs \cos vs}{\sin vs \sinh cs}
\end{equation}
exactly like equation (\ref{del02})
In fact the matrix calculation is the short distance
type IIA calculation of a configuration of a many zero branes and a membrane
bounded to many zero-branes.

This only happens because  $N$ is large, so that from the
point of view of type IIA string theory there are almost infinite number
of zero-branes on the membrane, so its effect are almost swamped by the
zero-branes. 
However if we expand equation (\ref{bj20}), rather than
(\ref{bj20c}),  in the lightest open string
modes and calculate the long range potential from that, we will find
a very different answer.

What happens when $c$ is not small. This is the situation in which
we have few zero-branes bounded to the membrane. Then the string calculation
using only the open string lowest modes will not reproduce the long
distance potential, because we are back to the expression (\ref{bj20}).
However the M(atrix) calculation will still reproduce 
the correct long range force  as long as $b^2 \gg c$. So going to
the infinite momentum frame has helped us to identify the essential
variables.

We now examine the issue of the tachyonic instability \cite{bansus}
 present in the
type IIA theory. As noticed in \cite{gil2} the tachyonic instability
starts at a shorter scale when there are zero-branes on the two-brane.
In the matrix calculation from equation (\ref{norm02}) one can see that there
is a tachyonic instability at $b^2 < c$, in agreement 
with the above statement.   
From our calculation we can extract the wave function of the open string
stretched between the zero-brane and the membrane bounded to zero-brane
system. The wave functions are just $\Psi(x)= e^{-\frac{1}{2c}x^2} 
H_n (x/\sqrt{c})$
where $H_n (x)$ are just the Hermite polynomial. $H_0(x)=1$ gives the
wavefunction for the tachyon.

If
$c$ is very small compare to $v$ then from (\ref{norm02}) one can see
that there is going to be a large imaginary part when $b^2 < v$ this is
just the nucleation of open string between the branes.


\subsection{Membrane - anti-Membrane}
An anti membrane is just a membrane with its orientation reversed.  A membrane
parallel to an anti membrane stretched in the $X_8,X_9$ direction 
and separated
in the $X_7$ direction by a distance b is described by the  configuration,

\[
B_8=\left(
\begin{array}{cc}
P_1 &  0  \\ 
0 & P_2 
\end{array}
\right), \ \
B_9=\left(
\begin{array}{cc}
Q_1 &  0  \\ 
0 & -Q_2 
\end{array}
\right), \ \
B_7=\left(
\begin{array}{cc}
0 &  0  \\ 
0 & bI 
\end{array}
\right).
\]
Where $[Q_1,P_1]=ic$ and $[Q_2,P_2]=ic$.
To see that this represents a membrane anti-membrane configuration one 
can observe a few things. First each configuration by itself
breaks only half the supersymmetries in equation (\ref{susy}), but
together they break all the supersymmetries. Second using the
results of \cite{berdoug} one sees that the anti
 membrane has as a fermionic zero mode
the one with opposite chirality than the membrane, this then gives
the opposite sign for the Berry phase and hence the opposite charge. Given
the membrane configuration there is more that one 
configuration representing the anti-membrane 
(although they are all equivalent), for instance
one can just take the membrane and exchange $P$ with $Q$ and vice versa,
or multiply $P$ or $Q$ by $-1$. Multipling both by $-1$ we get back a membrane.

Inserting this background to equation (\ref{l3}) one can read
 off the mass squared 
terms for the bosons and the fermions. 
Let us start with the bosons. in the space of 
$(Y_1,\ldots ,Y_7,C)$ the matrix
is diagonalised with diagonal entries all being $2H$ and,
\begin{equation}
H=b^2+(P_1+P_2)^2+(Q_1+Q_2)^2
\end{equation}
In $A_{0}$ space we get $-2H$
In the space of $Y_8,Y_9$ there 
are also off diagonal elements $\pm 8ic$.
After diagonalization and subtracting the ghost contribution we end up 
in Euclidean space with:
six bosons with $M^2=2H$ one with $M^2=2H-8c$ and one
with $M^2=2H+8c$.

Turning now to the fermions one finds 
\begin{equation}
m_f=\gamma_8(P_1+P_2)+\gamma_9(Q_1+Q_2)-\gamma_7bI
\end{equation}
So $M_{f}^{2}=H-2icI\gamma_8\gamma_9+Ib^2$.
This gives eight fermions with $M_{f}^{2}=H+2c$ and eight fermions
with $M_{f}^{2}=H-2c$ .

We now investigate  the spectrum of $H$. $H$ acts on a Hilbert 
space of functions of two variables.
We are realizing $P$ as derivative operator and 
$Q$ as position operator. Lets label 
the two variables by $x,y$. 
then $H = -4c^2\partial^{2}_{x+y} +(x+y)^2$. so it will
have eigenvalues of a harmonic oscillator but also 
a large  number of functions with
eigenvalue $b^2$
as $Hf(x-y)=b^2 f(x-y)$. The eigenvalue of $H$ is
$H_{n_+}=b^2+2c(2n_{+}+1)$, and there is a degeneracy which we will label
as $N_{-}$. The degeneracy is expected to be $\sim N$.

Evaluating the determinants as before and defining 
$r^{2}_{n}=b^2+2c(2n+1)$,  the potential (not the phase
shift) is 

\begin{equation}
V_n=N_{-}\frac{8}{\sqrt{\pi}}
\int \frac{ds}{s}s^{-1/2}e^{-r_{n}^{2}s}\sinh^4 cs.
\label{v2a2}
\end{equation}
Summing over $n$ 

\begin{equation}
V=N_{-}\frac{8}{\sqrt{\pi}}
\int \frac{ds}{s}s^{-1/2}e^{-r_{n}^{2}s}\frac{\sinh^4 cs}{2\sinh 2cs}.
\label{pot2a2}
\end{equation}
The long range potential is then,

\begin{equation}
V=N_{-}\frac{(2c)^3
\Gamma(5/2)}{4\sqrt{\pi}b^{5}}
\label{2a2m}
\end{equation}

We now turn to the type IIA configuration that matches the above configuration.
It is just a two-brane and an anti two-brane with some magnetic field 
$F$ on them. One can compute the potential between them as a function
of the separation $b$ and the magnetic field $F$.
\begin{equation}
V_{IIA}= L^2\int \frac{ds}{s}\frac{ e^{-b^2 s}}{\sqrt{4\pi s}}
 B \times J,
\label{a2a2}
\end{equation}
\begin{eqnarray}
B& = &\frac{2iF}{4\pi^2}f_{1}^{-8}\frac{\Theta_1'(0)}
{\Theta_{1}(i\epsilon s)}. \nonumber \\
J& = &\frac{1}{2}\{ -f_{2}^{8}
\frac{\Theta_{2} (i\epsilon s)}{\Theta_{2}(0)}
+f_{3}^{8}\frac{\Theta_{3} (i\epsilon s)}{\Theta_{3}(0)} \nonumber\\ 
 & + & f_{4}^{8} 
\frac{\Theta_{4} (i \epsilon s)}{\Theta_{4}(0)}\}.
\label{bj2a2}
\end{eqnarray}
Here  $F=\tan \frac{\pi\epsilon}{2}$ and $L^2$ is the
volume of the two-brane. Notice that
for $F=0$ one just gets the static potential between
a two-brane and an anti two-brane, and of course we
do not expect to get the full answer just from the lightest open 
string mode. However if $F$ starts becoming
large then things change. As $F \rightarrow \infty$ we write
$\epsilon =1-c'$, inserting that to equation 
(\ref{bj2a2}) and using the transformation properties
of the theta functions we get

\begin{eqnarray}
B& = &\frac{2iF}{4\pi^2}f_{1}^{-8}\frac{\Theta_1'(0)}
{\Theta_{1}(i c' s)}, \nonumber  \\
J& = &\frac{1}{2}\{ -f_{2}^{8}
\frac{\Theta_{2} (i c' s)}{\Theta_{2}(0)}
+f_{3}^{8}\frac{\Theta_{3} (i c' s)}{\Theta_{3}(0)} \nonumber\\ 
 & - & f_{4}^{8} 
\frac{\Theta_{4} (i c' s)}{\Theta_{4}(0)}\}.
\label{bj2a2f}
\end{eqnarray}
Notice now the crucial sign change in the third term in $J$. 
Equation (\ref{bj2a2f}) now looks like two two-branes with some small
magnetic field on one of them \cite{gregut,bacpor}, 
a configuration which we expect to have
the long distance potential being reproduced by the light open string modes.
This is the ''magic`` of the infinite momentum frame.

Expanding (\ref{bj2a2f}) for large $b$ one finds the long
range potential
\begin{equation}
V_{IIA}=\frac{L^2(\pi c')^2 \Gamma{5/2}}{2\pi^{3/2}}b^{-5}
\label{2a2s}
\end{equation}
Since $F=\tan \frac{\pi\epsilon}{2}$ and
$F=1/c$ we see that $\pi c'=2c$.
Comparing equations (\ref{2a2m}) and (\ref{2a2s}), and using
$c=\frac{L^2}{2\pi N}$  one sees that they
agree if $N_{-}=2N$. 
We will see a consistency check on this.

If we expand equation (\ref{bj2a2f}) in the limit 
where only the light open string
modes contribute we find (with $a_2$ some coefficient)
\begin{equation}
B \times J =a_2 \frac{\sinh^{4} \frac{\pi c' s}{2}}{\sinh \pi c's}
\end {equation}
exactly like equation (\ref{pot2a2}). This is another example
how the M(atrix) calculation is just a short distance type IIA calculation.

Notice that from equation (\ref{v2a2}) there is a tachyonic instability for
$b^2 <2c$.


\subsection{Two moving membranes}
Here we will consider two moving parallel membranes with
 relative velocity $v$.
The background configuration is given by: 
\[
B_8=\left(
\begin{array}{cc}
P_1 &  0  \\ 
0 & -P_2 
\end{array}
\right),
B_9=\left(
\begin{array}{cc}
Q_1 &  0  \\ 
0 & -Q_2 
\end{array}
\right),
B_7=\left(
\begin{array}{cc}
0 &  0  \\ 
0 & bI 
\end{array}
\right),
B_1=\left(
\begin{array}{cc}
0 &  0  \\ 
0 & vtI
\end{array}
\right).
\]
By now the derivation of the mass matrix is clear. We will first consider
the situation where $[Q_1,P_1]=ic_1=ic_2=[Q_2,P_2]$. 
We look for the  mass matrix in Euclidean space. 
For the bosons we find: six bosons with $M^{2}=2H$, one with
$M^2=2H+4iv$ and one with $M^2=2H-4iv$. Where 
$H=(P_1-P_2)^2+(Q_1+Q_2)^2+I(b^2+ v^2t^2)$.
For the fermions one finds: eight with $M_{f}^{2}=H+iv$  and eight with
$M_{f}^{2}=H-iv$.
the spectrum of $H$ is actually continuous in this case as 
$[(P_1-P_2),(Q_1+Q_2)]=0$. Define $x=Q_1+Q_2$, $y=Q_1-Q_2$, then
$(P_1-P_2)^2$ is realized as $-4c^2\partial_{y}^2$ and $(P_1+P_2)^2$
is realized as $-4c^2\partial_{x}^2$. Now $2dQ_1 d Q_2 \rightarrow dxdy$,
$2dP_1 d P_2 \rightarrow dqdk$, and the correct phase space
normalization has to include of course the $\frac{1}{2\pi}$.
The spectrum of $H$ is $H_{xk}=b^2+v^2t^2+4c^2k^2+x^2$. 
define:
\begin{equation}
r_{xk}=b^2+4c^2k^2+x^2
\end{equation}

the phase shift is then
\begin{equation}
\delta=8 N_{-}\int dx\frac{dk}{2\pi}
\int\frac{ds}{s}\frac{e^{-r_{xk}^{2}}}
{\sin sv}\sin^4 (sv/2).
\label{dn22}
\end{equation}
This give a potential
\begin{equation}
V=N_{-}\frac{v^4\Gamma(5/2)}{8c\sqrt{\pi}}b^{-5}
\label{22m}
\end{equation}
Where the integral over $x,k$ was approximated by taking its limits
to be $(-\infty, \infty)$.

What is the expected long range result from type IIA string theory?
The configuration from the string theory point of view is that
of two parallel moving membranes with a large world volume
magnetic field on each one but which is the same (remember $c_1=c_2$).
It is well known that the one loop amplitude in this case is just
the same as if there is no magnetic field, other than a multiple
coefficient of the form $(1+F^2)$ which
is just the Born-Infeld action \cite{acny,gregut,bacpor}.
In our case $F=\frac{1}{c}=\frac{2\pi N}{L^2}$, 
Then the potential between two relatively moving two-branes is \cite{bac,gil1}
\begin{equation}
V_{IIA}= \frac{L^2 v^4\Gamma(5/2)}{8c^2(\pi)^{3/2}b^{-5}}
\label{22s}
\end{equation}
Equations (\ref{22m}) and (\ref{22s}) agree if we use the degeneracy
that was used in the membrane anti-membrane case namely $2N$. This
is an independent check on this degeneracy.

If we would have taken  in this section $c_1 \neq c_2$
 one would find a non zero force 
even at zero velocity and there would be a tachyonic instability at short
distances  $b^2 < (c_1-c_2)$.


\subsection{Two orthogonal moving membranes}
Two orthogonal membranes is a supersmmetric configuration, but with only a 
quarter of the supersuumetries unbroken, thus we expect a $v^2$ term in the
potential.
the background for this configuration is
\[
B_8=\left(
\begin{array}{cc}
P_1 &  0  \\ 
0 &  0
\end{array}
\right),
B_9=\left(
\begin{array}{cc}
Q_1 &  0  \\ 
0 &  0 
\end{array}
\right),
B_7=\left(
\begin{array}{cc}
0 &  0  \\ 
0 & bI 
\end{array}
\right).
\]
\[
B_1=\left(
\begin{array}{cc}
0 &  0  \\ 
0 & vtI
\end{array}
\right),
B_5=\left(
\begin{array}{cc}
0  &  0  \\ 
0 &  P_2
\end{array}
\right),
B_9=\left(
\begin{array}{cc}
0 &  0  \\ 
0 &  Q_2 
\end{array}
\right).
\]
Performing the same calculations as before we find the spectrum of the bosons 
and fermions in Euclidean space.
 Define $H=P_{1}^{2}+Q_{1}^{2}+P_{2}^{2}+Q_{2}^{2}+b^2+v^2t^2$.
For the complex bosons there are two with $M^{2}=2H$, one with
$M^{2}=2H+4iv$ one with $M^{2}=2H-4iv$, one with $M^{2}=2H+4c_1$,
one with $M^{2}=2H-4c_1$ one with $M^{2}=2H+4c_2$ and one with
$M^{2}=2H-4c_2$. Remember that actually one has twice the number of real
bosons with half the mass squared.

For the fermions we have: two with $M_{f}^{2}=H+c_1+c_2+iv$, two with
$M_{f}^{2}=H+c_1+c_2-iv$, two with $M_{f}^{2}=H-c_1-c_2+iv$, two with
 $M_{f}^{2}=H-c_1-c_2-iv$, two with $M_{f}^{2}=H+c_1-c_2+iv$, two
with $M_{f}^{2}=H-c_1+c_2-iv$, two with $M_{f}^{2}=H+c_1-c_2-iv$, and
two with $M_{f}^{2}=H-c_1+c_2+iv$. 

For $c_1=c_2=0$ this is the
result  for two gravitons, and if $c_1=0\neq c_2$ then this is the 
graviton membrane scattering.

Let us first take $c_1=c_2=c$. The spectrum of $H$ is 
$H_{n_{1}n_{2}}=b^2+v^2 t^2 +c(2n_1+2n_2+2)$.
 then the phase shift becomes
\begin{equation}
\delta=\int \frac{ds}{s}e^{-b^{2}s}
\frac{2+2\cos 2vs +4 \cosh 2cs -4\cos vs -4 \cos vs \cosh2cs}
{8\sinh^2 cs \sin vs}
\end{equation}
Notice that for $v=0$ this vanishes as expected.
For large $b$ we can expand the integrand in powers of $s$,
this gives 

\begin{equation}
V=-\frac{(c^2v^2+\frac{1}{4}v^4)\Gamma(3/2)}{2c^2\sqrt{\pi}}b^{-3}.
\label{2or2m}
\end{equation}

The string theory configuration is just two orthogonal two-branes
with equal magnetic field on them. The phase shift is then 
given by the same expression for a zero-brane scattered off the
$(4-2-2-0)$ bound state in \cite{gil2}. let 
$\epsilon=\frac{1}{2}-c'$, 
the phase shift takes the form ($\tanh \pi \nu =v$)
\begin{equation}
\delta_{IIA}=\frac{1}{2\pi}\int \frac{ds}{s} e^{-b^2 s}
 B \times J.
\label{amp4220}
\end{equation}
\begin{eqnarray}
B& = -&\frac{1}{2}f_{1}^{-4}\Theta^{-2}_{1}(ic's)
\frac{\Theta_{1}' (0)}{\Theta_{1}(\nu t)}. \nonumber\\
J& = &\{ -f_{2}^{4}
\frac{\Theta_{2} (\nu s)}{\Theta_{2}(0)}
\Theta_{2}^{2}(ic's)+f_{3}^{4}\Theta_{3}^{2}(ic' s)
\frac{\Theta_{3} (\nu s)}{\Theta_{3}(0)} \nonumber \\
 & - & f_{4}^{4} 
\frac{\Theta_{4} (\nu s)}{\Theta_{4}(0)}\Theta_{4}^{2}(ic's)\}.
\label{bj2or2}
\end{eqnarray}
the long range potential  is 
\begin{equation}
V_{IIA}=-\Gamma(3/2)\frac{2(1-\cosh \pi \nu )\cos^2 \pi c' +\sinh^2 \pi \nu}
{2\sqrt{\pi} \sin^2 \pi c'} b^{-3}
\label{2or2s}
\end{equation}
Expanding to lowest order in $c'$ and $v$, one finds exactly the result of
(\ref{2or2m}), with $\pi c'=c$.

Again the expansion of the string result to the lowest modes of the 
open string will give exactly the M(atrix) result.

When $c_1 \neq c_2$ then there is a non zero force even at $v=0$ and when
$b$ is small enough there is a tachyonic instability.
Notice that there is a tachyonic instability when any two objects
have different $c$'s, that is when any two objects have different
eleven dimensional velocities, very similar to string nucleation
 when two brane are relatively moving \cite{bac,bacpor}.

\section{Conclusions}

In this paper we have calculated the potentials between various configurations
of 2-branes and gravitons by using the M(atrix) technique of \cite{bfss}.
We observed that the large boost made the configurations almost supersymmetric,
and allowed the short distance answers obtained from Yang-Mills theory
to reproduce also long range potentials, in accordance with one of the
hopes of the conjecture in \cite{bfss}. This fact is a more general 
version of
the fact that in a boosted frame transverse velocities become small and
nonrelativistic, which also brings us closer to a supersymmetric 
configuration.

One would, at the end, like to have potentials for branes that are not boosted,
and thus are not in supersymmetric configurations. This requires us to relate
the potentials obtained for the boosted branes to the potentials of
the branes without the boost. In the case of the 2-brane - anti-2-brane system
we saw in section 2 what form such a relation would take. The kinematics of
boosts in compact directions turned out to be somewhat different from the
naive kinematics that would be expected for boosts in noncompact directions.
It would be useful to have a more precise understanding of the kinematics in
the compact case.

Following \cite{grt} we have noted that 
the 2-brane bound to many 0-branes can be described as a theory of
the Yang-Mills for the 0-branes with a topological twist around the  
cycles where the two-brane is wrapped. This twist can be interpreted as
a unit of magnetic field  on the two dimensional space,
thus characterised by the first chern class of the bundle in which the
0-brane variables take values. In this same spirit we expect that
the 4-brane of type IIA theory, bound to many
0-branes,  could be described by a similar treatment of zero branes moving on
a 4-dimensional space with twists around the cycles corresponding to the
presence of an instanton on the 4-dimesional space.
 We hope to study the interactions of four-branes with two-branes
and gravitons in the future, using ideas similar to the ones
used here for the interactions between two branes and gravitons.

\begin{center}
{\bf Acknowledgements}
\end{center}
We would like to thank C. Callan, M.R. Douglas, O. Ganor, D. Kabat, 
I. Klebanov, S. Ramgoolam, S. Shenkar and W. Taylor for helpful
discussions. 
 

\end{document}